\documentclass[prd, aps, twocolumn, reprint, superscriptaddress, floatfix, amsmath, amssymb]{revtex4-2}

\usepackage{graphicx} 
\usepackage{dcolumn}  
\usepackage{bm}       
\usepackage{hyperref} 
\usepackage{slashed}  

\usepackage{xcolor}

\newcommand{\vb}[1]{\mathbf{#1}}
\newcommand{\vh}[1]{\hat{\mathbf{#1}}}

\begin{document}


\title{Low-mass dark sector searches with deuteron photodisintegration}

\author{Cornelis J.G. Mommers}
\email{cmommers@uni-mainz.de}
\author{Marc Vanderhaeghen}
\affiliation{Institut für Kernphysik and PRISMA${}^+$ Cluster of Excellence, \\ Johannes Gutenberg-Universität, D-55099 Mainz, Germany}

\date{\today}

\begin{abstract}

Recent years have seen much activity in searches for dark-sector messenger particles in the 10--100 MeV mass range, especially in view of a potential new light boson conjectured by the ATOMKI collaboration, X17.
Under the assumption that the messenger particle has definite parity and either zero or unit spin, quite stringent bounds already exist on its coupling to electrons and protons. 
Equally stringent bounds on the neutron coupling do not exist yet, but are nonetheless desirable.
We explore how measurements of deuteron photodisintegration with a quasi-free neutron can yield bounds on the neutron coupling, and compute projections for a potential measurement at the low-energy high-intensity electron scattering experiment MAGIX@MESA.
The projected bounds are found to be competitive for an axial-vector or pseudoscalar scenario, but not for a vector or scalar scenario.
\end{abstract}

\maketitle


\section{Introduction}\label{sec:intro} 

In recent years, the main paradigm for dark-matter searches has been shifting.  
With heavier candidate particles, such as WIMPs, either excluded or experimentally out of reach \cite{XENON:2018voc, ParticleDataGroup:2022pth}, a vigorous effort is presently underway to search for light dark-sector particles in the MeV--GeV mass range \cite{Beacham:2019nyx, Agrawal:2021dbo}. 
In this context a new boson $X$ is often posited to be a messenger particle connecting the known visible sector to a hidden dark sector \cite{ArkaniHamed:2008hhe, Essig:2013lka, Alexander:2016aln, Lanfranchi:2020crw}.

The mass of the mediator $X$ determines its decay. It may either decay invisibly into particles of the hidden dark sector, or it may decay visibly into particles of the Standard Model. 
Herein we deal with visible decays of the mediator $X$ to an electron-positron pair, where one typically searches for $X$ by directing a high-intensity, $\mathcal{O}$(100 MeV--few GeV) electron beam at a fixed gas-jet target \cite{Freytsis:2009bh, Essig:2010xa, Beranek:2013yqa}.
If present, a potential signal will appear as a narrow resonance on top of a large, but well-known QED background. 

Using these techniques robust exclusion limits on the coupling of $X$ to an electron-positron pair have been established, with reactions such as $e^- p \to e^-p (X \to e^+ e^-)$, or $e^- Z \to e^- Z (X \to e^+ e^-)$, where $Z$ denotes a heavy target nucleus \cite{Essig:2013lka}. 
Likewise, results from the NA48/2 experiment, using $\pi^0 \to \gamma(X \to e^+ e^-)$, have put strict bounds on the coupling of $X$ to the proton \cite{NA482:2015wmo}. 
And, under the assumption that $X$ is a pseudoscalar, results from the SINDRUM collaboration \cite{SINDRUM:1986klz}, which measured $\pi^+ \to e^+ \nu_e (X \to e^+ e^-)$, limit the isovector nucleon coupling of $X$.

Unlike the effective electron and proton couplings of $X$, the effective neutron coupling is not as stringently tested in the region $m_X \sim 10$--100 MeV$/c^2$. 
The reason is simple: in the lab, no free, high-density neutron target exists.
Any extractions of the neutron coupling have to be made by proxy, which is why the current strongest constraints come from neutron star phenomenology \cite{ParticleDataGroup:2022pth, Rrapaj:2015wgs, Berryman:2022zic} or isotope shift spectroscopy \cite{Berengut:2020itu, Rehbehn:2023khk}.
Unfortunately, these bounds fall short of covering the MeV-to-GeV mass range of interest. 

Nevertheless, for several reasons it is desirable to have better limits on the neutron coupling in this region as well. First, taken with the existing constraints on the proton coupling, constraints on the neutron coupling can be directly translated into limits on the quark couplings of $X$, i.e. $g_u = (2 g_p - g_n) / 3$ and $g_d = (2 g_n - g_p) / 3$.
In turn, these limits directly tell us what new physics models are permissible. As a specific example consider the dark photon which, through kinetic mixing with an ordinary photon, necessarily couples to the quarks in a manner proportional to their charges, $g_u = 2 g_p / 3$ and $g_d = - g_p / 3$, implying $g_n = 0$ \cite{Holdom:1985ag}.
Any deviation from this value already rules out the dark photon scenario. 
Another more immediate reason is connected to the potential discovery of a narrow resonance around 17 MeV by the ATOMKI collaboration---dubbed X17---in the $e^+ e^-$ decay spectrum of excited states of $^8$Be, $^{12}$C and $^4$He \cite{Krasznahorkay:2015iga, Krasznahorkay:2021joi, Krasznahorkay:2022pxs}. 
Limits on the coupling of X17 to the nucleon, which can be derived individually from each of the observed nuclear decays, appear to be in tension with each other \cite{Barducci:2022lqd, Denton:2023gat, Hostert:2023tkg}. 
Better constraints on the neutron coupling would help resolve this apparent conflict.

The problem, which involves determining the $X$ coupling to the neutron, is mirrored in the program to measure the neutron polarizabilities \cite{Griesshammer:2012we, Lensky:2015awa}. 
There, a successful strategy is to measure deuteron photodisintegration under specific kinematic conditions where the bound neutron behaves as though it were free \cite{A2CollaborationatMAMI:2021vfy}. 
One can then access the polarizabilities indirectly, using the `quasi-free' bound neutron \cite{Levchuk:1994ij}.
The purpose of this paper is to explore the use of a similar approach, a measurement of $\gamma d \to e^+ e^- pn$ in quasi-free neutron kinematics, to extract limits on the neutron coupling of $X$.

To make our analysis concrete, our work will be centered around the forthcoming experiment at MAGIX@MESA \cite{Doria:2019sux}, though it may be readily adapted to meet the constraints provided by other electron scattering facilities. 
MESA is an electron accelerator (presently under construction) that will be capable of producing a low-energy, high-intensity electron beam (up to 105 MeV in its energy-recovering mode).
With a gas-jet target, luminosities around $10^{35}$ cm$^{-2}\ $s$^{-1}$ will be reached \cite{A1:2021njh}.
The MAGIX experiment consists of two high-resolution electron-positron spectrometers with a relative momentum resolution of $\delta p / p < 10^{-4}$ and an angular resolution of $\delta \theta < 3$ mrad \cite{Doria:2019sux, A1:2021njh}. 
Accordingly, the expected electron-positron invariant-mass resolution is $\delta m_{ee} \approx 0.1$ MeV$/c^2$ or better \cite{Essig:2010xa}.
Note that earlier dark-photon searches have already achieved a resolution of 0.2 MeV$/c^2$ \cite{Merkel:2014avp, Backens:2021qkv}.

In this paper we will investigate which limits on the neutron coupling to a dark-sector messenger particle $X$ can be obtained with the MAGIX@MESA setup using $\gamma d \to e^+ e^- pn$. Our paper is structured as follows: in Sec.~\ref{sec:kin} we start by specifying the kinematics of the $\gamma d \to e^+ e^- pn$ process and derive an expression for the differential cross section. 
In Sec.~\ref{sec:PWIA} we subsequently describe the plane-wave impulse approximation, and outline the kinematic regime in which the scattering on a quasi-free neutron is valid. 
In Sec.~\ref{sec:BG_signal} we specify the QED background and $X$-boson signal, respectively.
In Sec.~\ref{sec:reach} we derive projections for the reach at MAGIX@MESA, and in Sec.~\ref{sec:X17} we apply our framework to the specific case of the conjectured X17.
We conclude with a brief outlook in Sec.~\ref{sec:outlook}. 

\section{Kinematics and cross section of $\gamma d \to e^+ e^- pn$}\label{sec:kin}

We begin our calculation with the kinematics of the $\gamma d \to e^+ e^- pn$ process. 
We work within the rest frame of the deuteron and set up our coordinate system such that the $z$-axis is along the direction of the incoming photon's three-momentum, and the neutron and photon three-momenta span the $x$-$z$ plane.
We have,
\begin{align*}
    \gamma(q, \lambda) \, d(p_d, M) &\to e^+(p_+, s_+) \, e^-(p_-, s_-) \\
    &\qquad \quad p(p_p, s_p) \, n(p_n, s_n),
\end{align*}
where $\lambda$ is the polarization of the incoming photon, $s_\pm$ are the $e^\pm$ helicities, and $M$, $s_p$ and $s_n$ are the deuteron, proton and neutron spin projection along the $z$-axis. 
In the \textit{lab} frame, the four-momenta are given by
\begin{align*}
    p_d^\mu &= (m_d,      \, 0, \, 0, \, 0), \quad
    q^\mu  = (E_\gamma, \, 0, \, 0, \, E_\gamma) ,\\ 
    p_i^\mu &= (E_i, \, |\vb{p}_i| \vh{p}_i), \quad \text{for } i = \pm, n, \\    
    p_p^\mu &= (E_p, \, \vb{p}_p),
\end{align*}
where
\begin{align}
    \vh{p}_i &= (\sin\theta_i \cos\phi_i, \, \sin\theta_i \sin\phi_i, \, \cos\theta_i), \text{ for } i = \pm, n, \\
    \vb{p}_p &= \vb{q} - \vb{p}_- - \vb{p}_+ - \vb{p}_n.
\end{align}
We denote the masses of the nucleons, leptons and deuteron as  $m_N$, $m_e$ and $m_d$, respectively. 
It is also useful to introduce the four-momentum $q' = p_+ + p_-$ and the squared invariant mass of the dilepton system, $q'^2 = m_{ee}^2$.

We choose our kinematic variables as $E_\gamma$, $|\vb{p}_\pm|$, the polar angles $\theta_\pm$ and $\theta_n$, and the azimuthal angles $\phi_\pm$ and $\phi_n$. 
It follows that the magnitude of the neutron three-momentum is given by,
\begin{equation}\label{eq:kin}
    |\vb{p}_n| = \frac{1}{2a} \left[ b \pm \sqrt{b^2 - 4ac} \right],
\end{equation}
where
\begin{align}
    a &= (E_\gamma + m_d - q'^0)^2 - |\vb{q} - \vb{q}'|^2 \cos^2\theta_{n\gamma\gamma} > 0  ,\\
    b &= (q + p_d - q')^2 |\vb{q} - \vb{q}'| \cos\theta_{n\gamma\gamma} ,\\
    c &= m_N^2 (E_\gamma + m_d - q'^0)^2 - \frac{1}{4}(q + p_d - q')^4 > 0 ,
\end{align}
with
\begin{equation}
    | \vb{q} - \vb{q}' | \cos\theta_{n\gamma\gamma} = \vh{p}_n \cdot ( \vb{q} - \vb{q}' ).
\end{equation}
For our kinematic regime of interest the positive solution of Eq.~\eqref{eq:kin} corresponds to physical values of the neutron three-momentum. 

Lastly, the differential cross section is given by
\begin{equation} \label{eq:cross_section}
    \frac{ \mathrm{d} \sigma }{\mathrm{d}\Pi} = K \langle | \mathcal{M} |^2 \rangle,
\end{equation}
with
\begin{align}
    K &= \frac{1}{ 64 ( 2\pi )^{8} m_d E_\gamma }  \frac{ | \vb{p}_+ |^2 | \vb{p}_- |^2 }{E_+ E_-} \nonumber \\
    &\qquad \times  \frac{ | \vb{p}_n |^2 }{| \vb{p}_n | \left( m_d + E_\gamma - q'^0 \right) - E_n | \vb{q} - \vb{q}' | \cos\theta_{n\gamma\gamma} }, \label{eq:K}
\end{align}
where $\mathrm{d}\Pi$ is shorthand for $ \mathrm{d} |\vb{p}_+| \mathrm{d} | \vb{p}_- | \mathrm{d} \Omega_-  \mathrm{d} \Omega_+  \mathrm{d} \Omega_n$. Furthermore, in Eq.~\eqref{eq:cross_section} $\lvert \mathcal{M} \rvert^2$ is the squared Feynman amplitude and $\langle \, \cdot \, \rangle$ denotes the average over initial helicities or spin projections and sum over final helicities or spin projections.

\section{The plane-wave impulse approximation}\label{sec:PWIA}

\begin{figure}[t]
    \centering
    \includegraphics[width=8.6cm]{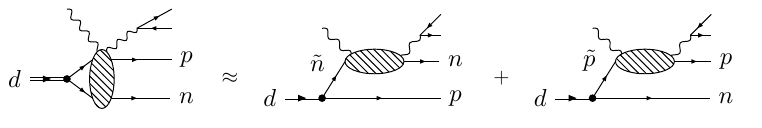}
    \caption{In the plane-wave impulse approximation we can factorize the process $\gamma d \to e^+ e^- pn$ into a process on a quasi-free neutron, $\tilde{n}$ and a process on a quasi-free proton, $\tilde{p}$.}
    \label{fig:PWIA}
\end{figure}

As a first step, to calculate the matrix element $\mathcal{M}$ we work within the plane-wave impulse approximation (PWIA) \cite{Levchuk:1994ij}.
In this approximation, shown in Fig.~\ref{fig:PWIA}, we can separate the process $\gamma d \to e^+ e^- pn$ into two parts: a part where the proton is a spectator and a part where the neutron is a spectator. 
By doing so we disregard meson exchange currents and final state interactions. 
However, in our kinematic regime of interest the meson exchange currents are estimated to give corrections of approximately 5\%, meaning they can be safely neglected.
Likewise, for a first approximation, the final state interactions can be omitted (see Fig.~5 and its discussion in Ref.~\cite{Levchuk:1994ij}).

As we are interested in the low-energy regime ($E_\gamma \sim 100$ MeV) where relativistic corrections are expected to be small, we use a non-relativistic framework to implement the PWIA. 
We start by inserting a complete set of two-particle states \footnote{We do not need to include a label for isospin, as a potential isospin factor of $1/\sqrt{2}$ is already included in the CB-Bonn wave function \cite{Machleidt:2000ge}.},
\begin{widetext}
\begin{align}
    \mathcal{M}_\text{IA}(\gamma d \to e^+ e^- pn) &= \sum_{s_1, s_2} \int \frac{\mathrm{d}^3 \vb{p}}{(2\pi)^3} \, \langle e^+ e^-, \, \vb{p}_p \, s_p, \, \vb{p}_n \, s_n \, \vert \, \mathcal{M}_\text{IA} \, \vert \, \gamma, \, \tfrac{1}{2} \vb{p}_d + \vb{p} \, s_1, \, \tfrac{1}{2} \vb{p}_d - \vb{p} \, s_2 \rangle \nonumber \\
    &\qquad \qquad \times \frac{(2m_d)^{1/2}}{(2E_{\tilde{p}})^{1/2} (2E_{\tilde{n}})^{1/2}} \, {}_\text{NR}\langle \vb{p}; \, s_1 \, s_2 \, \vert \, d(1, M) \rangle_\text{NR}, \label{eq:PWIA_NR}
\end{align}
\end{widetext}
where `NR' indicates the baryon states are normalized non-relativistically and no subscript refers to covariant normalization,
\begin{align*}
    \vert \vb{p}, \, s \rangle &= (2 E_p)^{1/2} \vert \vb{p}, \, s \rangle_\text{NR}, \\
    \langle \vb{p}, \, s \, \vert \, \vb{p}', \, s' \rangle &= (2 E_p) (2\pi)^3 \delta^{(3)}(\vb{p} - \vb{p}') \delta_{ss'}.
\end{align*}

In the second line of Eq.~\eqref{eq:PWIA_NR} we can identify the components of the relative deuteron wave function in momentum space, $\tilde{\Psi}^M_{s_1 s_2}(\vb{p}) = {}_\text{NR} \langle \mathbf{p};  \, s_1 \, s_2 \, \vert \, d(1, M) \rangle_\text{NR}$. 
They are given by
\begin{align}
    \tilde{\Psi}^M_{s_1 s_2}(\vb{p}) &= (2\pi)^{3/2} \bigg[ \tilde{\psi}_0(p) Y^0_0(\vh{p}) \langle \tfrac{1}{2} \, s_1; \, \tfrac{1}{2} \, s_2 \, \vert \, 1 \, M \rangle  \nonumber \\
    &- \tilde{\psi}_2(p) \sum_{M_s} Y^{M - M_s}_2(\vh{p}) \langle 1 \, M_s; \, 2 \, M - M_s \, \vert \, 1 \, M \rangle \nonumber \\
    &\quad \times \langle \tfrac{1}{2} \, s_1; \, \tfrac{1}{2} \, s_2 \, \vert \, 1 \, M_s \rangle \bigg] ,
\end{align}
where $p = |\mathbf{p}|$, $\langle j_1 \, m_1; \, j_2 \, m_2 \, \vert \, j \, m \rangle$ are Clebsch-Gordan coefficients and $Y^m_l$ are the spherical harmonics.
For our numerical estimates we use the CD-Bonn parametrization \cite{Machleidt:2000ge} for the s- and d-wave functions. Note that we use a different convention for the Fourier transform as compared to Ref.~\cite{Machleidt:2000ge}, which gives rise to the different factors of $2\pi$.
Also note that the relative sign between the s- and d-wave functions is absorbed into the CD-Bonn parametrization of Ref.~\cite{Machleidt:2000ge}. That is, $\tilde{\psi}_0(p) = \tilde{\psi}^\text{Bonn}_0(p)$ and $\tilde{\psi}_2(p) = -\tilde{\psi}^\text{Bonn}_2(p)$. 
The wave function is normalized as
\begin{equation}
    1 = \int \frac{\mathrm{d}^3 \vb{p}}{(2\pi)^3} | \tilde{\Psi}^M (\vb{p}) |^2 = \int_0^\infty \mathrm{d}p \, p^2 \left[ \tilde{\psi}_0^2(p) + \tilde{\psi}_2^2(p) \right].
\end{equation}

In this work we will restrict ourselves to kinematics inside the neutron quasi-free peak (NQFP) \cite{Levchuk:1994ij}. 
Qualitatively, this means we only consider configurations in which the bound proton is a spectator, i.e. has low momentum. 
As a consequence the incoming photon primarily scatters off the bound neutron $\tilde{n}$, and contributions to the cross section from processes involving the bound proton are negligible.   

A more quantitative definition of the NQFP may be derived using the relative and total momenta of the bound neutron and bound proton. 
In the lab frame the total momentum is just the deuteron momentum, $\vb{p}_d = \vb{0}$, and the relative momentum is the momentum of the outgoing proton, $\vb{p} = \vb{p}_p = - \vb{p}_{\tilde{n}}$.
Near the NQFP $| \vb{p} | \to 0$, meaning we approach the singularity of the deuteron wave function,
\begin{equation*}
    \frac{1}{m_d - E_{p} - E_{\tilde{n}} } = \frac{1}{2 m_N - \Delta - 2 E_{p} },
\end{equation*}
where $\Delta \approx 2.2$ MeV is the deuteron binding energy.
Close to this singularity, $\tilde{n}$ is `nearly on shell', and so we define the NQFP by \cite{Levchuk:1994ij}
\begin{equation}\label{eq:NQFP}
    E_p \lesssim m_N + \frac{\Delta}{2} \implies |\vb{p}_p| \lesssim \sqrt{m_N \Delta} \approx 45.7 \text{ MeV/c}.
\end{equation}

Expanding $\tilde{\Psi}^M$ in Eq.~\eqref{eq:PWIA_NR} in the NQFP region gives
\begin{equation}\label{eq:PWIA_NQFP}
    \mathcal{M}_\text{IA}(\gamma d \to e^+ e^- pn) \approx \mathcal{M}^n_{\text{IA}}(\gamma d \to e^+ e^- pn),
\end{equation}
with
\begin{align}
    &\mathcal{M}^n_{\text{IA}}(\gamma d \to e^+ e^- pn) = (2 m_d)^{1/2} \left(\frac{E_p}{E_{\tilde{n}}}\right)^{1/2} \nonumber \\
    &\quad \times  \sum_{s_{\tilde{n}}} \tilde{\Psi}^{M}_{s_p s_{\tilde{n}}} \left( \vb{p}_p \right) \mathcal{M}\left( \gamma \, \tilde{n} \to e^+ \, e^- \, n \right),
\end{align}
where the quasi-free neutron has momentum and spin projection $-\vb{p}_p$ and $s_{\tilde{n}}$, respectively. 
We have explicitly checked the size of the proton contribution to the cross section and found it to be negligible in the kinematic region corresponding with Eq.~\eqref{eq:NQFP}, validating Eq.~\eqref{eq:PWIA_NQFP}. 
Note that in the NQFP kinematic region of interest here any off-shell effects are small, and can be safely neglected.

\section{QED background processes and signal processes} \label{sec:BG_signal}

Having related the amplitude of $\gamma d \to e^+ e^- pn$ to the amplitude of $\gamma \tilde{n} \to e^+ e^- n$, we turn to the calculation of $\mathcal{M}(\gamma \tilde{n} \to e^+ e^- n)$.
This calculation is composed of two steps, the calculation of the QED background amplitude and the subsequent computation of the signal amplitude. 

\subsection{QED background amplitudes}

\begin{figure}[t]
    \centering
    \includegraphics[width= 8.6cm]{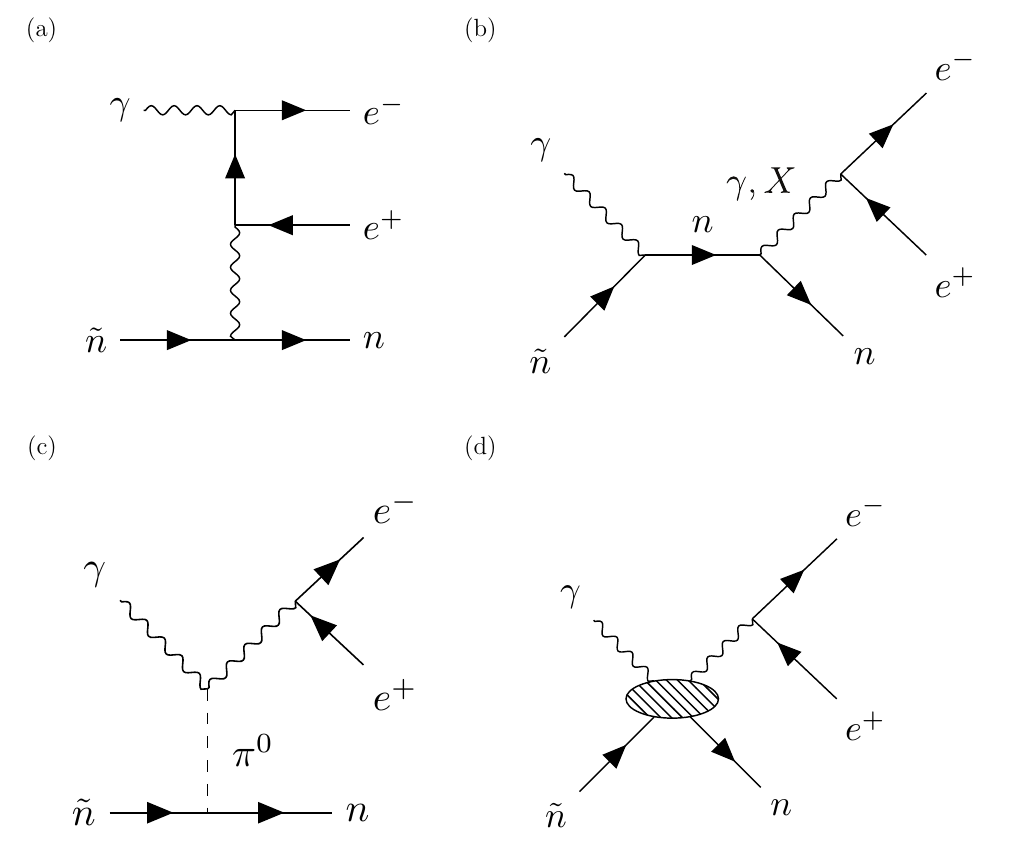}
    \caption{Below the pion production threshold the reaction $\gamma d \to e^+ e^- pn$ is described by the Bethe-Heitler process (a), the Born process (b), $\pi^0$ $t$-channel exchange (c) and non-Born neutron scalar polarizability contributions (d). Crossed diagrams for the Bethe-Heitler and Born processes are not shown explicitly. All diagrams are embedded in the quasi-free neutron blob in Fig.~\ref{fig:PWIA}.}
    \label{fig:QED_diag}
\end{figure}

For photon energies around 100 MeV, i.e. below the pion production threshold, the QED background may be parameterized by a combination of the Bethe-Heitler process and Compton scattering process. 
The latter is described by the Born process, the $\pi^0$ $t$-channel exchange process and the non-Born contributions parameterized by the neutron electric ($\alpha_E$) and magnetic ($\beta_M$) polarizabilities. 
These are all shown in Fig.~\ref{fig:QED_diag}, where we have omitted the crossed diagrams for the Bethe-Heitler and Born contributions.
All diagrams in Fig.~\ref{fig:QED_diag} are embedded in the quasi-free neutron blob of Fig.~\ref{fig:PWIA}.
Technical details of the calculation of the QED background have been relegated to Appendix~\ref{app:QED_BG} and a comparison of our calculated QED background and previous calculations is given in Appendix~\ref{app:QED_compare}.

\subsection{Signal amplitudes}
The $X$ signal process is identical to Fig.~\ref{fig:QED_diag}b, with $X$ replacing the virtual photon.
In principle, $X$ may also contribute via the Bethe-Heitler process. 
However, to extract limits on the neutron coupling from $\gamma d \to e^+ e^- pn$, we are only interested in processes where $X$ is on resonance (that is, $m_{ee}^2 = m_X^2$). 
Consequently, any Bethe-Heitler contributions stemming from virtual $X$ exchange are negligibly small and may be safely omitted.

We assume that $X$ is either scalar, pseudoscalar, vector or axial vector ($S$, $P$, $V$ or $A$, respectively) and that it couples to the fermions $f$, where $f$ may be a neutron, proton or electron ($f = n$, $p$ or $e$, respectively).
We can then generically parameterize its interaction using the effective interactions,
\begin{equation}
    \mathcal{L}_i = \sum_{f} g^i_f \bar{f} \Gamma_i f \cdot X, \quad \text{for } i = S,\, P,\, V,\, A, \label{eq:Lag}
\end{equation}
where $\Gamma_S = 1$, $\Gamma_P = i\gamma_5$, $\Gamma^\mu_V = \gamma^\mu$ and $\Gamma^\mu_A = \gamma^\mu \gamma_5$. The dot in Eq.~\eqref{eq:Lag} indicates all Lorentz indices, should they be present, are to be contracted. 
In this approach we remain agnostic to any models describing $X$;
constraints on the effective couplings $g^i_f$ can always be matched to constraints on couplings from a UV-complete model by integrating out the corresponding degrees of freedom of the microscopic theory.

Using Eq.~\eqref{eq:Lag} we find that the signal amplitude is given by
\begin{equation}\label{eq:X_amp}
    \mathcal{M}_i = \frac{B_i}{m^2_{ee} - m_X^2 + i m_X \Gamma_X}, \quad \text{for } i = S,\, P,\, V,\, A,
\end{equation}
where $B_i = \varepsilon(\vb{q}, \lambda) \cdot H_i \cdot C_i \cdot L_i$, with 
\begin{align}
    C^{(\mu \nu)}_i &= -e g^i_e g^i_n [ \delta_{Si} + \delta_{Pi} -  g^{\mu\nu} \delta_{Vi} \nonumber \\
    &\qquad - \left( g^{\mu \nu} - q'^\mu q'^\nu / m_X^2 \right) \delta_{Ai} ], \\
    L^{(\mu)}_i &= \bar{u}(p_-, s_-) \Gamma_i^{(\mu)} v(p_+, s_+), \\
    H^{\mu(\nu)}_i &= \bar{u}(p_n, s_n) \bigg\{ \Gamma_i^{(\nu)} \frac{(\slashed{p}_{\tilde{n}} + \slashed{q} + m_N)}{(p_{\tilde{n}} + q)^2 - m_N^2} \Gamma^\mu_q \nonumber\\
    &\qquad + \Gamma^\mu_q \frac{(\slashed{p}_{\tilde{n}} - \slashed{q}' + m_N)}{(p_{\tilde{n}} - q')^2 - m_N^2} \Gamma_i^{(\nu)} \bigg\} u(p_{\tilde{n}}, s_{\tilde{n}}).
\end{align}
The photon nucleon vertex $\Gamma^\mu_q$, is described in Eq.~\eqref{eq:NNg}, with $e > 0$.

\subsection{Signal averaged over a single invariant lepton-mass bin}

As can be seen from Eq.~\eqref{eq:X_amp}, the final cross section around the NQFP with $m_{ee} = m_X$ will generally depend on the coupling of $X$ to the neutron and electron, as well as the (narrow) width of $X$, $\Gamma_X$.
In an experiment we would not be able to set $m_{ee} = m_X$ exactly---instead, we will collect all events falling in $\left[m_{ee} - \delta m_{ee}/2, \, m_{ee} + \delta m_{ee} /2 \right]$.
If $\delta m_{ee}$ is sufficiently narrow, $\sim \mathcal{O}(0.1$ MeV$/c^2)$, the QED background will remain nearly constant across this bin and, given that $\Gamma_X \ll \delta m_{ee}$, this is also the only bin in which the signal will reside.

To calculate the signal averaged over a single bin we start by fixing $\delta m_{ee} = 0.1$ MeV$/c^2$ (bin widths achievable by MAGIX@MESA).  
Moreover, on resonance we may safely neglect any interference terms between the QED background and the $X$-boson signal due to the narrowness of the resonance,
\begin{equation}
    \frac{ \mathrm{d} \sigma }{\mathrm{d}\Pi} \approx \left( \frac{ \mathrm{d} \sigma }{\mathrm{d}\Pi} \right)_\text{QED} + \left( \overline{\frac{ \mathrm{d} \sigma_i }{\mathrm{d}\Pi}} \right)_{X}, \quad \text{for } i = S, P, V, A,
\end{equation}
where
\begin{equation}\label{eq:sig_avg}
    \left( \overline{\frac{ \mathrm{d} \sigma_i }{\mathrm{d}\Pi}} \right)_X := \frac{1}{\delta m_{ee}} \int^{m_X + \delta m_{ee} / 2}_{m_X - \delta m_{ee} / 2} \mathrm{d}m_{ee} \, \left( \frac{ \mathrm{d} \sigma_i }{\mathrm{d}\Pi} \right)_X,
\end{equation}
is the signal averaged over a bin. 
We assume that $m_X \gg \Gamma_X$, so that in Eq.~\eqref{eq:sig_avg} we may replace,
\begin{equation}
    \bigg\lvert \frac{1}{m^2_{ee} - m_X^2 + i m_X \Gamma_X} \bigg\rvert^2 \to \frac{\pi}{m_X \Gamma_X} \delta(m^2_{ee} - m_X^2). 
\end{equation}
This gives
\begin{equation}
    \left(\overline{\frac{ \mathrm{d} \sigma_i }{\mathrm{d}\Pi}}\right)_X 
    = \frac{1}{\delta m_{ee}}\frac{\pi}{2 m_X^2 \Gamma_X}
    \big( K \langle \lvert B_i \rvert^2 \rangle \big)\big\vert_{m^2_{ee} \, = \, m_X^2},    
\end{equation}
where $K$ is given in Eq.~\eqref{eq:K} and $B_i$ is given in Eq.~\eqref{eq:X_amp}.

We can rewrite $\Gamma_X$ using 
\begin{equation*}
    \mathcal{B}(X \to e^+ e^-) = \frac{\Gamma(X \to e^+ e^-)}{\Gamma_X}.
\end{equation*}
A short calculation gives
\begin{align}
    \Gamma_i(X \to e^+ e^-) &= \frac{\left( g_e^i \right)^2}{\pi} \big[ \left( \delta_{Pi} + \delta_{Vi} \right) p_\text{cm} \nonumber \\
    &\qquad + \left( \delta_{Si} + \delta_{Ai} \right) p^3_\text{cm} / m_X^2 \big] \beta_i,
\end{align}
where $\beta_S = 1$, $\beta_P = 1/4$, $\beta_V = (1 + 2m_e^2 / m_X^2)/6$ and $\beta_A = 2/3$, with $p_\text{cm} = \sqrt{m_X^2 - 4m_e^2}/2$.

Using the above, $\overline{\mathrm{d}\sigma / \mathrm{d}\Pi}$ no longer directly depends on $\Gamma_X$ and $g_e^i$, and only depends on the electronic branching fraction, which appears as an overall scaling factor. Throughout this work we set $\mathcal{B}(X \to e^+ e^-) = 1$. If the branching fraction is not equal to unity, any result can be easily rescaled by a simple multiplication.

\section{Neutron-coupling reach for MAGIX@MESA}\label{sec:reach}

Next, we present projections for the reach of the neutron couplings for an experiment at MAGIX@MESA.
We consider a signal significant if it is $n_\sigma$ standard deviations over the QED background.
Then, the reach at the $n_\sigma$ confidence level is given by \cite{Beranek:2013yqa}
\begin{equation}\label{eq:reach_explicit}
    |g_n^\text{reach}| = \left[ \frac{\sigma_\text{QED}}{\sigma_X\vert_{g_n = 1}} \frac{n_\sigma}{\sqrt{L \times  \sigma_\text{QED}}} \right]^{1/2},
\end{equation}
where $\sigma_i$ are the integrated signal and background cross sections and we denote the integrated luminosity by $L = \int \mathrm{d}t \, \mathcal{L}$, with $\mathcal{L}$ the luminosity.
Equation~\eqref{eq:reach_explicit} implies an approximate scaling of the reach with the bin width like $\sqrt{\delta m_{ee}}$. 

\begin{figure*}[t]
    \centering    
    \includegraphics[width=17.4cm]{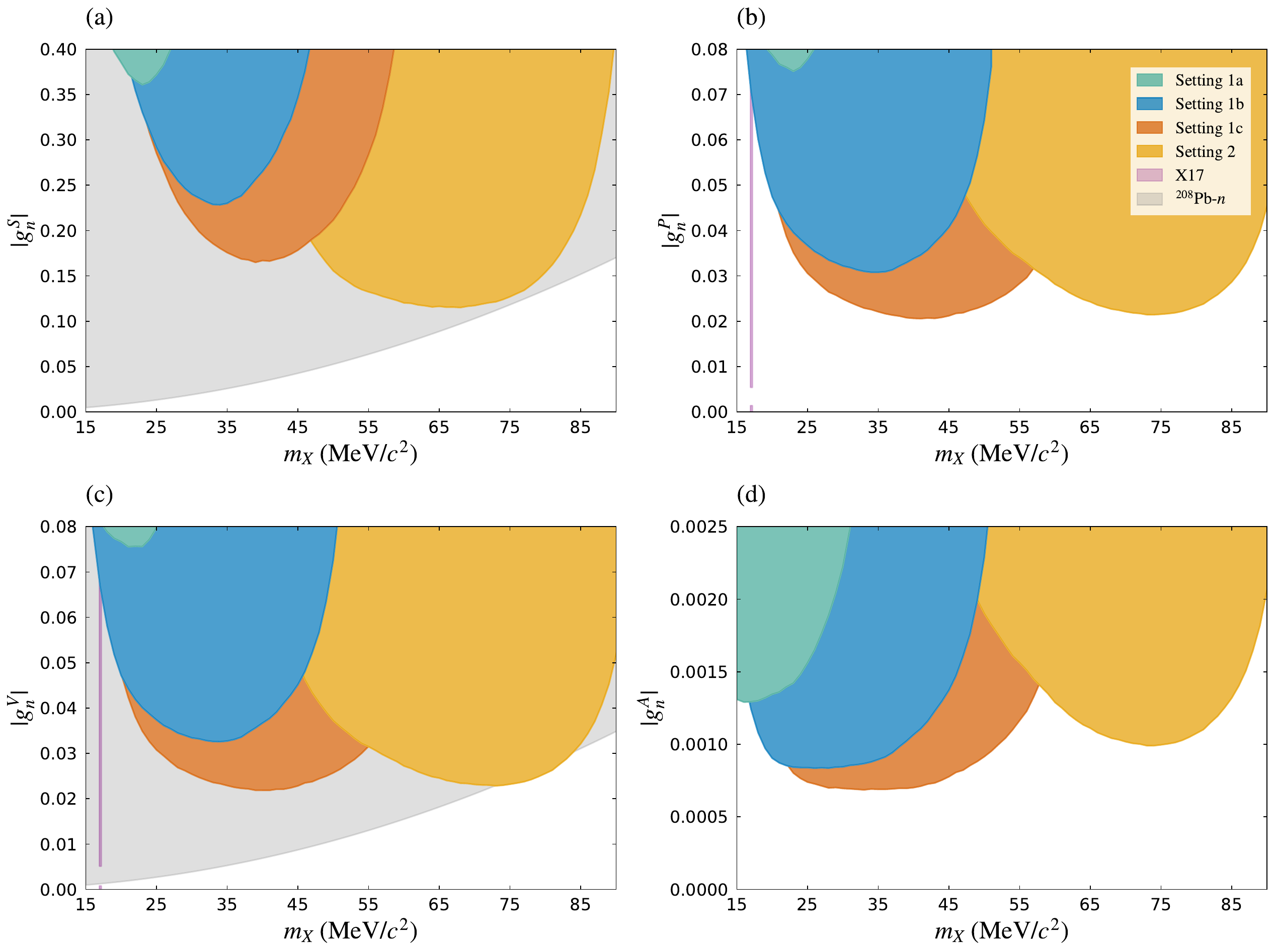}
    \caption{$2\sigma$ reach plots for the effective neutron coupling assuming a scalar (a), pseudoscalar (b), vector (c) or axial-vector (d) boson, where any signal in the shaded yellow, orange, blue or cyan regions would be detectable. In all plots the bin width $\delta m_{ee} = 0.1$ MeV$/c^2$ and the integrated luminosity $L = 7.2 \times 10^8$ nb$^{-1}$. Results may be easily rescaled using Eq.~\ref{eq:reach_explicit}. The detector settings are given in Table~\ref{tab:detector}.
    Existing bounds from neutron-lead scattering \cite{Barbieri:1975xy, Barducci:2022lqd} are shown in gray, while bounds from the ATOMKI experiments at 17 MeV/$c^2$ are shown in purple. The latter are discussed in detail in Sec.~\ref{sec:X17} and Fig.~\ref{fig:X17}. We note that for all scenarios there exist several bounds on either the difference of the neutron coupling and an (unbounded) proton coupling, and many bounds on parameters other than the neutron coupling. Both cannot be directly shown in the above figure. For further discussion regarding constraints on parameters other than the neutron coupling, see Sec.~\ref{sec:reach} and, for instance, Refs.~\cite{Barducci:2022lqd, Alves:2017avw, Kozaczuk:2016nma, ParticleDataGroup:2022pth, Hostert:2023tkg, Denton:2023gat, Kahn:2016vjr} and references therein.}
    \label{fig:reach}
\end{figure*}

The range over which we can integrate the signal and background cross sections is restricted by: a) the quasi-free neutron condition~\eqref{eq:NQFP}, b) the requirement that $m_{ee}$ falls into the correct bin (with width $\delta m_{ee} = 0.1$ MeV$/c^2$), and c) geometric constraints from MAGIX@MESA \cite{A1:2021njh}. 
The latter limit $|\theta_\pm| \leq 165\text{\textdegree}$ and $|\theta_n| \geq 5\text{\textdegree}$.
Even with these constraints the accessible phase space is much larger than can be realistically measured in a single run. 
To that end, we consider several more restrictive detector settings.
Across all settings we fix
\begin{align*}
    |\vb{p}_\pm| &\in [10.0, \, 100.0] \text{ MeV}/c, \\ 
    \phi_+ &\in [150.0\text{\textdegree}, \, 210.0\text{\textdegree}],\\
    \phi_- &\in [-30.0\text{\textdegree}, \, 30.0\text{\textdegree}], \\
    \theta_n &\in [-30.0\text{\textdegree}, \, -5.0\text{\textdegree}] \cup [5.0\text{\textdegree}, \, 30.0\text{\textdegree}],
\end{align*}
and vary the other kinematic variables as outlined in Table \ref{tab:detector}. 

In all settings we only consider backward lepton kinematics because the signal-to-background ratio is considerably more favorable as compared to forward lepton kinematics. 
This is because in forward lepton kinematics the Bethe-Heitler part of the QED background increases, thereby significantly worsening the signal-to-background ratio.

We stress that the integration limits we provide are estimates.
Naturally, they are to be refined once the parameters for the MAGIX@MESA setup are established.
However, projections in Ref.~\cite{A1:2021njh} suggest our estimates are likely to be realistic.
Furthermore, should there be any significant changes then our analysis can easily be repeated. 
Finally, note that in practice the lepton spectrometers are restricted to in-plane kinematics only \cite{A1:2021njh}. 
To cover a larger range of $\phi_\pm$ one can equivalently measure the neutron at an out-of-plane angle.  

\addtolength{\tabcolsep}{4pt}    
\begin{table}[tb]
\caption{The detector settings considered in this work, consistent with the projected values for MAGIX@MESA \cite{A1:2021njh}. The $m_{ee}$ range follows immediately from $m_{ee}^2 = (p_+ + p_-)^2$, where we have rounded the range to the nearest multiple of $5$ MeV$/c^2$. Note that limits similar to setting 2 but with a lower photon energy are fully covered by settings 1b and 1c.} 
\label{tab:detector}
\centering
    \begin{tabular}{llll}
    \hline
    \hline
    Setting & $E_\gamma$ (MeV) & $\theta_\pm$ (deg) & $m_{ee}$ range (MeV/$c^2$) \\ \hline
    1a & 55  & 135.0--165.0 & 10.0--35.0     \\
    1b & 85  & 135.0--165.0 & 15.0--50.0       \\
    1c & 105 & 135.0--165.0 & 20.0--65.0    \\
    2  & 105 & 105.0--135.0 & 45.0--90.0    \\
    \hline
    \hline
    \end{tabular}%
\end{table}
\addtolength{\tabcolsep}{-4pt}

We perform the integration using Monte Carlo integration with 10,000 integration points, at which results converge to within 1--2\% of each other, which we deem to be sufficiently accurate for our purposes here.
We use rejection sampling to implement the NQFP condition \eqref{eq:NQFP} and to ensure all events fall within the desired $m_{ee}$ bin \cite{SHIRLEY199280}.

The reach, assuming $n_\sigma = 2$, is shown in Fig.~\ref{fig:reach}, for a scalar (a), pseudoscalar (b), vector (c) and axial-vector (d) $X$ scenario.
To calculate the reach we have taken the projected luminosity of MAGIX@MESA's gas-jet target, $\mathcal{L} = 10^{35}$ cm$^{-2}$ s$^{-1}$ \cite{A1:2021njh}, together with a beam time of $2000$ h, giving an integrated luminosity $L = 7.2 \times 10^8$ nb$^{-1}$.
Of course, results in Fig.~\ref{fig:reach} can easily be rescaled using Eq.~\eqref{eq:reach_explicit} to account for different luminosities. 

It is evident that a higher $E_\gamma$ correlates with an improved reach, and the best (lowest) reach is obtained in the 20--85 MeV$/c^2$ range with $E_\gamma = 105$ MeV.
To access masses $\lesssim 20$ MeV$/c^2$ at $E_\gamma = 105$ MeV requires measurements in which the leptonic polar angles $\theta_\pm$ are further backwards than is geometrically feasible for the MAGIX@MESA setup.
To access the lower mass region, one instead has to lower $E_\gamma$.
As can be seen by comparing settings 1a--c, this unfortunately worsens the reach for the lower masses. 

\begin{figure}[t]
    \centering    
    \includegraphics[width=8.6cm]{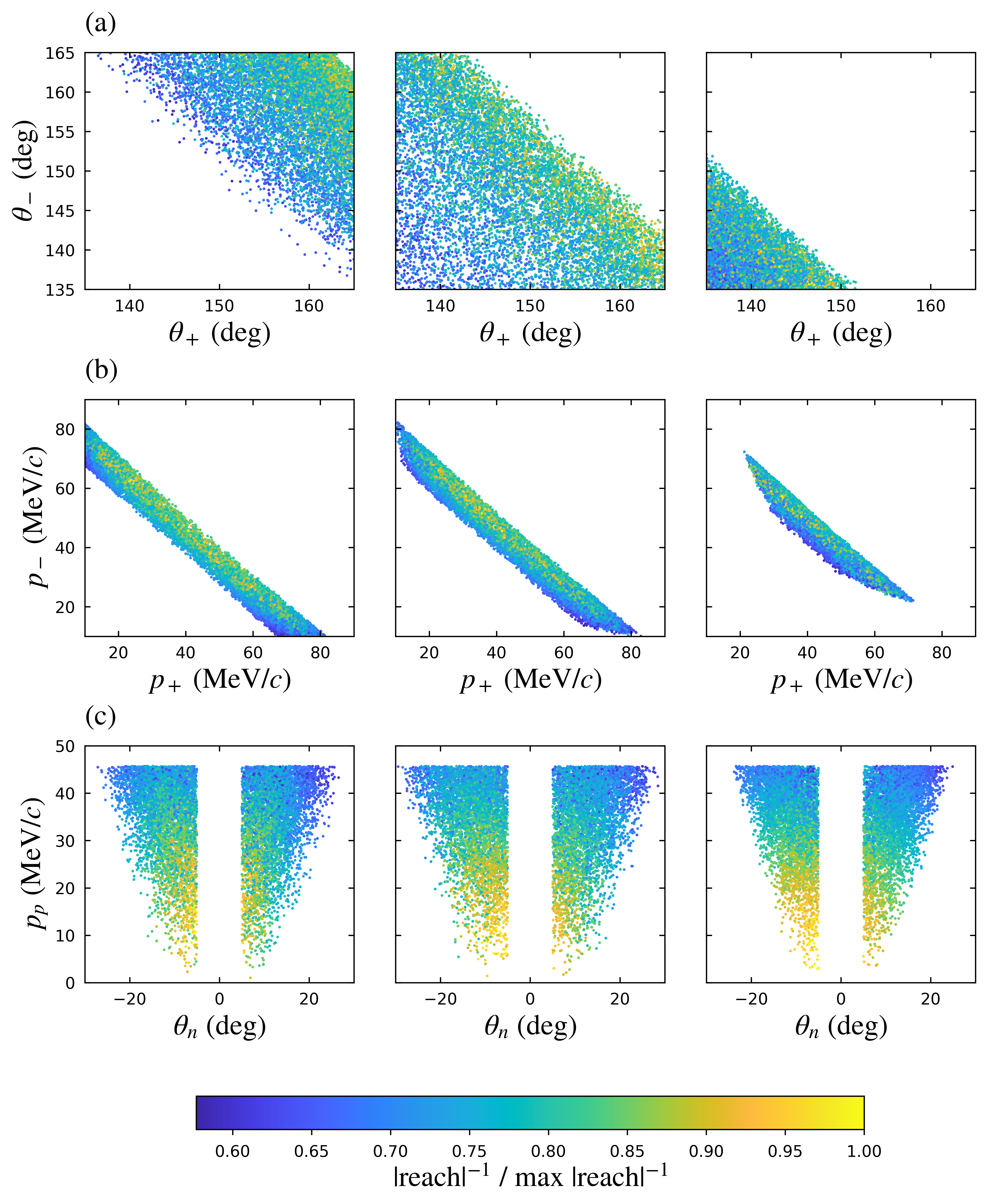}
    \caption{Projections of the phase space from 10,000 uniformly sampled points specified by detector setting 1c (see table \ref{tab:detector}) for $m_{ee} = 25$, 40 and 55 MeV$/c^2$ (left to right, respectively) and $\delta m_{ee} = 0.1$ MeV$/c^2$, assuming a vector-like scenario. The color code shows the inverse reach normalized to its maximum value of the considered sample (see Eq.~\eqref{eq:reach_explicit}).}
    \label{fig:density}
\end{figure}

To further explore the relation between the reach and our kinematic variables, in Fig.~\ref{fig:density} we show several projections of the phase space (sampled 10,000 times) assuming a vector-like scenario for setting 1c, with $\delta m_{ee} = 0.1$ MeV$/c^2$ and $m_{ee} = 25$, 40 and 55 MeV$/c^2$. In Fig.~\ref{fig:density} we have, at each kinematic point, taken the inverse of Eq.~\eqref{eq:reach_explicit} using the differential cross section instead of the integrated cross section, yielding $|\text{reach}|^{-1}$. In addition, we have normalized each point to the maximum value of $|\text{reach}|^{-1}$ out of the 10,000 sampled points, denoted by $\max |\text{reach}|^{-1}$. 
We have checked, and broadly our conclusions also hold for other parity assignments, bin sizes and settings.

In Fig.~\ref{fig:density}a one sees that a smaller $m_{ee}$ corresponds to more backward leptonic polar angles $\theta_\pm$. 
This is because $m_{ee}$ is proportional to the relative angle between three-momentum of the lepton pair, meaning if $m_{ee}$ is small, this relative angle has to be small as well. 
It is also evident that backward lepton kinematics give a better reach than forward kinematics.
The reason for this is, as previously mentioned, that in the forward regime the Bethe-Heitler contribution is enhanced, thus washing away the signal.

In Fig.~\ref{fig:density}b we have plotted the lepton momenta $|\mathbf{p}_\pm|$. Here one sees that higher momenta generally correlate to a better reach. In passing, let us also mention that projections in $\phi_\pm$ (not shown) are fairly evenly distributed.

Finally, in Fig.~\ref{fig:density}c we have plotted the proton momentum $|\vb{p}_p|$ against the neutron angle $\theta_n$.
For all $m_{ee}$ a clear pattern is visible: a lower protonic three-momentum corresponds to a better reach, which is not surprising. 
From Eq.~\eqref{eq:reach_explicit} we see that the reach is determined by the signal-to-background ratio as well as the absolute size of the integrated QED cross section.
The absolute value of the proton three-momenta enters the cross section in $\tilde{\Psi}^M(\vb{p}_p)$.
The most prominent contribution there, the deuteron s-wave function, is peaked at small momenta values. 
Thus, one obtains a larger cross section for small $|\vb{p}_p|$, which is exactly reflected in Fig.~\ref{fig:density}.

As an aside, we remark that a cut on small $|\vb{p}_p|$ would not be advantageous.
We have checked that the gain in signal-to-background (the first fraction in Eq.~\eqref{eq:reach_explicit}) is smaller than the loss in phase space (the second fraction in Eq.~\eqref{eq:reach_explicit}) after placing the cut.
Therefore, the net result is a worse reach.
We discuss this further in Appendix~\ref{app:reach_int_range}.

In Fig.~\ref{fig:diff_cs} we have plotted the differential cross section of the background and signal as a function of $\theta_n$. Here we assumed an axial-vector $X$ scenario with $m_X = 17$ MeV$/c^2$, though similar conclusions hold for other parity assignments and masses as well. We fixed $E_\gamma = 105$ MeV, $\delta m_{ee} = 0.1$ MeV$/c^2$ and the other kinematic variables to,
\begin{align*}
    |\vb{p}_+| &= 65.7 \text{ MeV}/c^2, \quad \theta_+ = 165.0 \text{\textdegree}, \quad \phi_+ = -180.0 \text{\textdegree}, \\
    |\vb{p}_-| &= 20.1 \text{ MeV}/c^2, \quad \theta_- = 165.0 \text{\textdegree}, \quad \phi_- = 0.0 \text{\textdegree}, \\
    \phi_n &= 0.0 \text{\textdegree}.
\end{align*}
The QED background is shown in red. In the above kinematics $m_{ee} = m_X$, leading to the enhanced signal shown in blue. For the blue curve, $g_n^A = 7 \times 10^{-5}$, while the light blue bands correspond to the range $g_n^A = (5$--$9) \times 10^{-5}$. The angular dependence shown in Fig.~\ref{fig:diff_cs} is typical for kinematics in the NQFP. The maximum of the peaked cross section corresponds approximately to the angle $\theta_n$ at which $|\vb{p}_p|$ reaches its minimum, assuming all other kinematic variables are kept fixed.

\begin{figure}[t]
    \centering    
    \includegraphics[width=8.6cm]{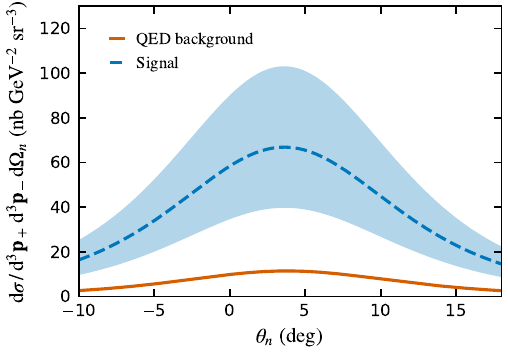}
    \caption{The differential signal cross section (blue) assuming an axial-vector scenario and the differential QED background cross sections (red) with uncertainties due to the neutron polarizabilities. We fix $E_\gamma = 105$ MeV, $m_{X} = 17$ MeV$/c^2$, $\delta m_{ee} = 0.1$ MeV$/c^2$ and $g^n_A = 7 \times 10^{-5}$. The bands correspond to the range $g_n^A = (5$--$9) \times 10^{-5}$. A further specification of the kinematics is given in the body of the text.}
    \label{fig:diff_cs}
\end{figure}

\begin{figure}[t]
    \centering    
    \includegraphics[width=8.6cm]{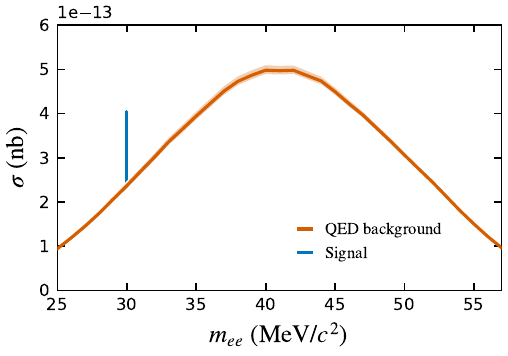}
    \caption{The integrated signal (blue) and QED background (red) with uncertainties due to the neutron polarizabilities for $m_{X} = 30$ MeV$/c^2$, $\delta m_{ee} = 0.1$ MeV$/c^2$ and $g^n_A = 10^{-5}$ using detector setting 1c (see Table~\ref{tab:detector}). A hypothetical axial-vector boson, $X$, would show up inside a single bin where $m_{ee} = m_X$.}
    \label{fig:signal}
\end{figure}

Lastly, in Fig.~\ref{fig:signal} we have plotted the integrated background and signal cross section for an axial-vector $X$ scenario as a function of $m_{ee}$, with $m_X = 30$ MeV$/c^2$ and $\delta m_{ee} = 0.1$ MeV$/c^2$ using detector setting 1c.
Analogous results hold for different values of $m_X$, the bin width, and different parity assignments.
The signal height is proportional to $|g^A_n|^2$;
for illustrative purposes only we have chosen $g^A_n = 10^{-5}$. 
In an experiment, the strongly-enhanced signal would clearly be visible inside a single bin. 

We may compare the obtained reach with some available limits in the literature. As can be seen in Fig.~\ref{fig:reach}a, for the scalar scenario limits from Ref.~\cite{Barbieri:1975xy},
\begin{equation}
    \lvert g_n^S \rvert \times (m_X \text{ MeV}/c^2)^{-2} \lesssim 2.1 \times 10^{-5},
\end{equation}
already exclude the accessible parameter space.

Similarly, for a vector-like $X$ we have constraints from $^{208}$Pb-$n$ scattering \cite{Barducci:2022lqd}, limiting 
\begin{equation}
    \lvert g_n^V \rvert \times \bigg \lvert \frac{126}{208} g_n^V + \frac{82}{208} g_p^V \bigg \rvert \times (m_X \text{ MeV}/c^2)^{-4} \lesssim 4.3 \times 10^{-10}.
\end{equation}
The intensive search program for the dark photon has yielded strong bounds on the proton coupling \cite{NA482:2015wmo}, which we may reinterpret as bounds on a more general vector $X$ \cite{Feng:2016ysn},
\begin{equation}
    |g^V_p| \lesssim 2.5 \times 10^{-4}. 
\end{equation}
Combining both constraints implies
\begin{equation}
    |g_n^V| \times (m_X \text{ MeV}/c^2)^{-2} \lesssim 4.3 \times 10^{-6},
\end{equation}
which also already excludes the accessible parameter space, as can be seen in Fig.~\ref{fig:reach}c.

For a pseudoscalar $X$ current limits on the nucleon coupling are given in terms of the isovector coupling \cite{Alves:2017avw},
\begin{equation}
    | g^P_p - g^P_n | \lesssim 1.2 \times 10^{-3}. 
\end{equation}
Such an inequality may be easily satisfied if the proton and neutron coupling are of a similar order, so the purely-neutron exclusion limits one gets from $\gamma d \to e^+ e^- pn$ become relevant again. 

A similar situation arises for an axial-vector $X$, where, out of all parity assignments, the couplings are least constrained. The strictest bounds comes from the KTeV anomaly \cite{KTeV:2006pwx, Barducci:2022lqd},
\begin{equation}\label{eq:KTeV}
    1.3 \times 10^{-10} \leq \frac{\left( g^A_p - g^A_n \right) g^A_e}{(m_X \text{ MeV}/c^2)^{2}} \leq 5.2 \times 10^{-10}, 
\end{equation}
where the upper and lower bounds are given in units of $\left(\text{MeV}/c^2 \right)^{-2}$.
The electron coupling can be estimated to be 
\begin{equation}
    g^A_e = \pm(1.52 \pm 0.31) \times 10^{-4},
\end{equation}
which gives two possible bands in which the neutron coupling may fall (see Fig.~\ref{fig:X17} in Sec.~\ref{sec:X17} and Fig.~5 of Refs.~\cite{Barducci:2022lqd, Hostert:2023tkg}).
Moreover, just as the pseudoscalar coupling, one can fine-tune the proton and neutron couplings to largely evade the bound on their difference. 
Therefore, bounds solely on the neutron coupling would be a very welcome complement to the constraints given above.

On the other hand, one should also consider recent results from the isotope-shift analysis of Xenon \cite{Rehbehn:2023khk} (and similar systems \cite{Berengut:2020itu, Leeb:1992qf}).
At low energies both a pseudoscalar and an axial-vector interaction are described by a Yukawa potential in position space \cite{Fadeev:2018rfl}. When included in isotope shift analyses, one obtains bounds on the product of the electron and neutron couplings. These bounds typically go up to 1 MeV$/c^2$ and are of $\mathcal{O}(10^{-11})$. If we extrapolate them to our energy range, we get an order-of-magnitude estimate $|g_e g_n| \lesssim \mathcal{O}(10^{-9})$. Current bounds on the electron coupling are $\mathcal{O}(10^{-4})$ \cite{Alves:2017avw}, which would mean the isotope-shift analyses would give limits on the neutron coupling of $\mathcal{O}(10^{-5})$, much stricter than the ones we get in Figs.~\ref{fig:reach}b and d. Of course, it remains an open question whether atomic methods can be improved enough to access the 10--100 MeV$/c^2$ mass range any time soon.

Next, we discuss how one may further improve the bounds in Fig.~\ref{fig:reach}.
A possible approach would be to decrease $\delta m_{ee}$. 
However, given that halving $\delta m_{ee}$ to 0.05 MeV$/c^2$ only gives an improvement of the reach with a factor $\sim 1/\sqrt{2}$, significant improvements from this area are unlikely.
As previously stated and demonstrated in Appendix~\ref{app:reach_int_range}, the reach can also be enhanced by increasing the measured phase space.
Given that the ranges given in Table~\ref{tab:detector} already represent (roughly) the largest ranges that would be realistically possible for MAGIX@MESA to measure, the only remaining variable is the integrated luminosity.
As we already consider a fairly long experiment given our choice of 2000 hours, we turn to the luminosity.
It is unlikely that $\mathcal{L}$ of the gas-jet target with a deuteron gas can be increased enough at MAGIX@MESA to make up the difference, as the luminosity enters as $\mathcal{L}^{-1/4}$ into the reach.
However, the gas-jet target does allow for different gases with more neutrons---for example Xenon, with around 80 neutrons. 
One naively expects the luminosity (and by extension the reach) to scale with the number of available neutrons.
To that end, interesting further research would be to investigate the coupling limits around the NQFP of a more neutron-rich system. 

In sum, deuteron photodisintegration gives competitive limits on the neutron coupling for a pseudoscalar or axial vector $X$, though it is likely that in the future methods such as isotope-shift spectroscopy will yield improved limits.
That being said, whereas other methods give limits on the product of $g_n$ with another coupling, an extraction using $\gamma d \to e^+ e^- pn$ gives limits directly on $g_n$. Moreover, it is expected that by replacing the deuteron with a more neutron-rich system the reach can be improved further.

\section{Further results related to the X17 anomaly}
\label{sec:X17}

As mentioned in the introduction, recent results of the ATOMKI collaboration have garnered significant theoretical and experimental interest; in a series of experiments \cite{Krasznahorkay:2015iga, Krasznahorkay:2021joi,Krasznahorkay:2022pxs, Krasznahorkay:2023sax} the collaboration claims to have found evidence of a new, light boson dubbed X17.

The ATOMKI collaboration looked at internal pair creation in decays of excited ${}^8$Be, ${}^4$He and, recently, ${}^{12}$C nuclei.
In all three cases, an anomalous bump was found in the distribution of the emitted electron-positron pair's relative angle, with a statistical significance consistently exceeding $6 \sigma$ (see Ref.~\cite{Alves:2023ree} for a review).
Working with members of the ATOMKI collaboration, researchers at VNU University of Science claim to have independently verified the original $^8$Be measurements \cite{Anh:2024req}. 

In the Standard Model, nuclear transitions in which an $e^+ e^-$ pair is emitted are mediated by electromagnetic interactions and are well understood. 
They are sensitive to new physics appearing at the MeV scale, and thus the ATOMKI collaboration attributes their anomaly to the as-of-yet unseen X17, with a reported averaged mass around $17.02(10)$ MeV$/c^2$ \cite{Krasznahorkay:2022pxs,Krasznahorkay:2021joi, Krasznahorkay:2015iga}. 
Assuming definite parity, the beryllium results indicate X17 can be a pseudoscalar, vector or axial-vector particle \cite{Krasznahorkay:2015ijz}, while the carbon results point to a scalar, vector or axial-vector particle \cite{Barducci:2022lqd}. 
Theoretically, based on the ${}^8$Be decays, models for X17 have been developed for the pseudoscalar, vector and axial-vector cases \cite{Feng:2016ysn,Ellwanger:2016wfe,Kozaczuk:2016nma,Alves:2017avw,Viviani:2021stx} that attempt to explain the ATOMKI anomalies while conforming to existing exclusion bounds. 
In particular, according to the vector model put forward by Feng \textit{et al}. \cite{Feng:2016jff,Feng:2016ysn} X17 additionally must be protophobic (couple weakly to protons) to meet existing bounds from the NA48/2 experiment \cite{NA482:2015wmo}. 
Experimentally, a global effort is underway to scrutinize the results of the ATOMKI anomaly, with new experiments at facilities such as CCPAC \cite{Azuelos:2022nbu}, MEG II at the PSI \cite{MEGII:2018kmf}, JLAB \cite{Dutta:2023ifr}, New JEDI \cite{Bastin:2023utm}, among many others \cite{Alves:2023ree}.

We investigate how the neutron-coupling limits from deuteron photodisintegration discussed above can be of added value in the current search for X17.
To estimate the coupling of X17 to the nucleon we compare models by Alves and Weiner \cite{Alves:2017avw} for the pseudoscalar case, by Feng \textit{et al}. \cite{Feng:2016ysn} for the vector case and by Kozaczuk \textit{et al}. \cite{Kozaczuk:2016nma} for the axial-vector case. 
To match their notation with ours we have 
\begin{align*}
    g^{(0)}_{XNN} &= (g^P_p + g^P_n) / 2, \quad g^{(1)}_{XNN} = (g^P_p - g^P_n) / 2,\\
    \varepsilon_f &= g^V_f / e , \\
    a_f &= g^A_f,
\end{align*}
for $f = p$, $n$ and $e$.

\begin{figure*}[t]
    \centering    
    \includegraphics[width=17.4cm]{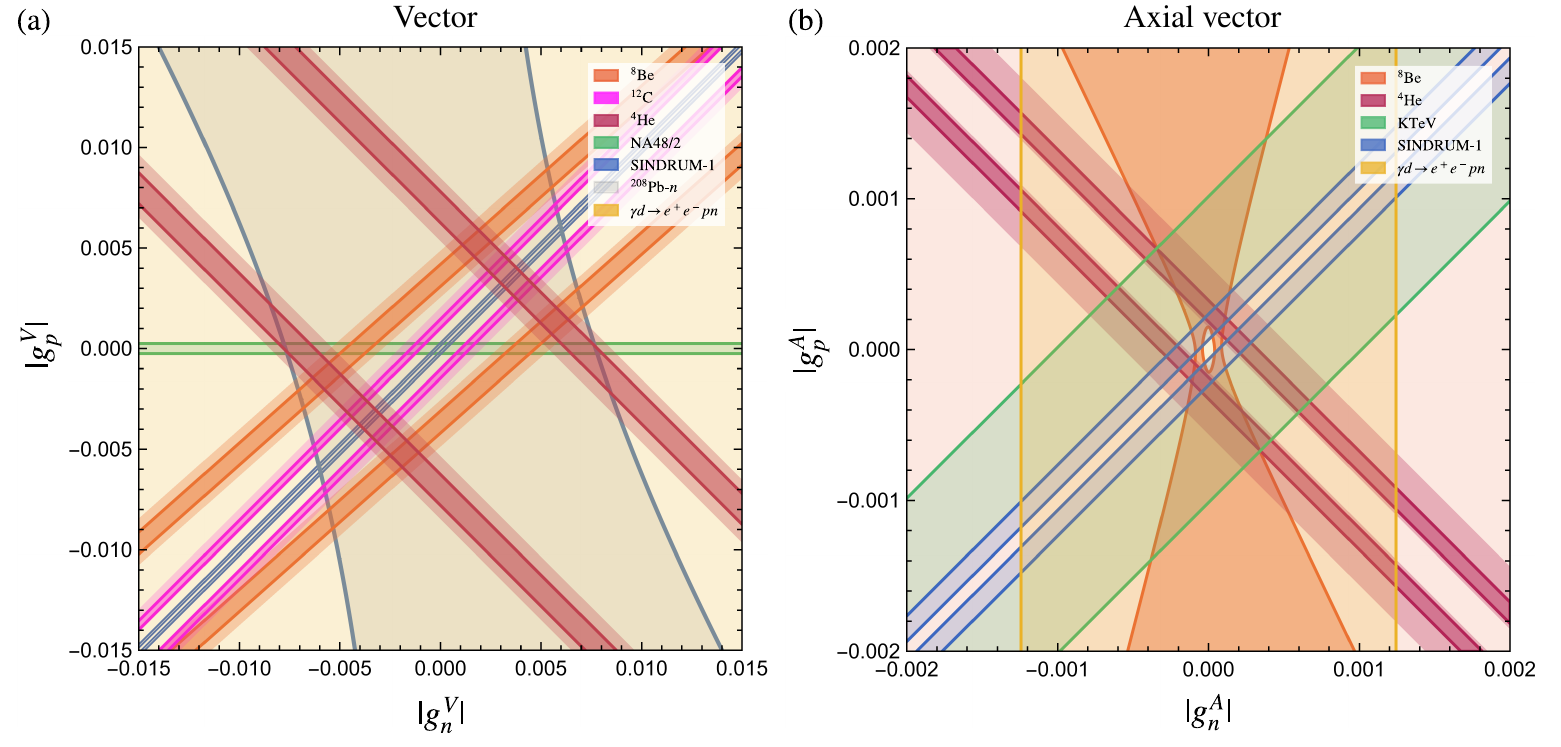}
    \caption{The $1\sigma$ (dark) and $2\sigma$ (light) allowed parameter space for a vector (a) or axial-vector (b) X17. Projections from $\gamma d \to e^+ e^- pn$ from Fig.~\ref{fig:reach} are shown in yellow. As can also be seen in Fig.~\ref{fig:reach}, for the vector case, this constraint is subdominant to other constraints.  
    The ${}^8$Be, ${}^4$He and ${}^{12}$C-derived constraints are given in orange, red and pink, respectively, where we use the isospin mixing angle $\theta_{1^+} = 0.35(8)$\textdegree{} for the $^8$Be states (see Sec.~\ref{sec:X17}). A different mixing angle rotates the $^8$Be bands around the origin.
    For the vector scenario additional constraints from NA48/2 \cite{NA482:2015wmo} and lead-neutron scattering \cite{Barbieri:1975xy} are shown in green and gray, respectively. Pion decay constraints \cite{Hostert:2023tkg} are shown in blue, where the upper and lower bands correspond to a positive and negative electron coupling, respectively.
    For the axial-vector scenario additional constraints from the KTeV anomaly \cite{KTeV:2006pwx} and pion decay \cite{Hostert:2023tkg} are shown in green and blue respectively, where the upper bands corresponds to a positive electron coupling and the lower bands to a negative electron coupling.}
    \label{fig:X17}
\end{figure*}

We can constrain X17's couplings by using the branching fractions of the ${}^8$Be and ${}^{12}$C decays reported by the ATOMKI collaboration \cite{Krasznahorkay:2015iga,Krasznahorkay:2022pxs},
\begin{align}
    \frac{\Gamma_X}{\Gamma_\gamma} \bigg\vert_{{}^8\text{Be}(18.15)} &= 6(1) \times 10^{-6}, \label{eq:Gbe}\\
    \frac{\Gamma_X}{\Gamma_\gamma} \bigg\vert_{{}^{12}\text{C}(17.23)} &= 3.6(3) \times 10^{-6}.
\end{align}
as well as from the branching fraction of $^4$He decays. The latter are more complicated with furthers details given in Refs.~\cite{Krasznahorkay:2021joi,Feng:2020mbt, Barducci:2022lqd}. Because the couplings of X17 have been extensively discussed in other works we shall only give a brief summary.
For more details we refer the reader to the relevant papers. 

The ${}^8$Be(18.15) state, which is predominately isoscalar, is isospin mixed with the ${}^8$Be(17.64) state, which is predominately isovector.
In our analysis we parameterize this isospin mixing with an isospin-mixing angle, $\theta_{1^+}$, and an isospin-breaking parameter, $\kappa$ \cite{Feng:2016ysn}. 
Following Ref.~\cite{Alves:2017avw}, we take $\theta_{1^+} = 0.35(8) \text{\textdegree}$, whence $\kappa = 0.681$ \cite{Feng:2016ysn}. 

For a pseudoscalar X17 scenario, results from the SINDRUM collaboration \cite{SINDRUM:1986klz} put a strong bound on the isovector coupling, 
\begin{equation}
    |g^{(1)}_{XNN}| \leq 0.6 \times 10^{-3}.
\end{equation}
By following the procedure described in Ref.~\cite{Alves:2017avw} we derive bounds on the isoscalar coupling from Eq.~\eqref{eq:Gbe}. 
This gives
\begin{equation*}
    |g^{(0)}_{XNN}| = (2.0 - 5.4) \times 10^{-3}, \quad |g^{(1)}_{XNN}| = (0.0 - 0.6) \times 10^{-3}.
\end{equation*}
Clearly, from Fig.~\ref{fig:reach}b our projected limits from deuteron photodisintegration for a pseudoscalar X17 are not competitive.

For a vector X17 scenario the constraint provided by the NA48/2 experiment \cite{NA482:2015wmo} leads to the protophobia condition (the green band in Fig.~\ref{fig:X17}a), 
\begin{equation}
    |\varepsilon_p| \leq 1.2 \times 10^{-3}.
\end{equation}
We derive the remaining neutron coupling from the ${}^8$Be data as outlined in Ref.~\cite{Feng:2016ysn}, and from the ${}^{12}$C data using \cite{Barducci:2022lqd}
\begin{equation}
    \frac{\Gamma_X}{\Gamma_\gamma} \bigg\vert_{{}^{12}\text{C}(17.23)} = \frac{k}{\Delta E} \left( 1 + \frac{m_X^2}{2 \Delta E^2} \right) | \varepsilon_p - \varepsilon_n |^2,
\end{equation}
where $k = \sqrt{\Delta E^2  - m_X^2}$. 
We find
\begin{align*}
    ^8\text{Be: } &|\varepsilon_n| = (1.1 - 1.7) \times 10^{-2} , \\
    ^{12}\text{C: } &|\varepsilon_n| = (2.6 - 5.3) \times 10^{-3}.
\end{align*}
The projected limits from $\gamma d \to e^+ e^- pn$, shown in Fig.~\ref{fig:reach}, would have to be improved by at least a factor of two to test these couplings. This is also visible in Fig.~\ref{fig:X17}a, where the $\gamma d \to e^+ e^- pn$ projections span the entire parameter space.

In passing, we remark that like Refs.~\cite{Barducci:2022lqd, Denton:2023gat, Hostert:2023tkg}, we find some tension between the carbon, beryllium and helium results for a vector-like X17.
The neutron couplings only overlap when the uncertainty of the ${}^8$Be results is increased to around 3$\sigma$ (see the orange, red and pink bands in Fig.~\ref{fig:X17}a). 
In view of this there is an ongoing discussion in the literature surrounding X17's couplings. 
There is a general consensus that, as it stands, one cannot explain the ${}^8$Be, ${}^4$He and ${}^{12}$C anomalies simultaneously without introducing some form of tension (see Refs.~\cite{Denton:2023gat,Hostert:2023tkg} for more discussion thereon. From pion decays, the blue bands in Fig.~\ref{fig:X17}a, one may obtain additional constraints which increase the tension even more). 
Our carbon-derived neutron couplings agree with Ref.~\cite{Denton:2023gat}, which would indicate that one should critically re-examine the beryllium data; a similar finding is echoed in Ref.~\cite{Hostert:2023tkg}. 
A possible resolution of the discrepancy would be if X17 simply is not a vector particle. 
If X17 exists, and if it were in truth an axial vector, then interpreting its axial couplings as vector couplings would, naturally, lead to erroneous results.

To derive couplings for the scenario of an axial-vector X17 we need its nuclear matrix elements.
For the beryllium transition we take the matrix elements as calculated in Ref.~\cite{Kozaczuk:2016nma},
\begin{align*}
    \langle {}^8\text{Be}(\text{g.s.}) \vert \vert \hat{\sigma}^{(p)} \vert \vert {}^8\text{Be}(18.15) \rangle &= (-0.38 \pm 2.19) \times 10^{-2}, \\
    \langle {}^8\text{Be}(\text{g.s.}) \vert \vert \hat{\sigma}^{(n)} \vert \vert {}^8\text{Be}(18.15) \rangle &= (-10 \pm 2.6) \times 10^{-2},
\end{align*}
which may be used in \cite{Barducci:2022lqd}
\begin{align}
    &\Gamma_X \big\vert_{^{8}\text{Be}(18.15)} = \nonumber \frac{k_X}{18\pi} \left[ 2 + (\Delta E / m_X)^2 \right] \nonumber \\
    &\quad \times \lvert \langle {}^8\text{Be}(\text{g.s.}) \vert \vert a_p \hat{\sigma}^{(p)} + a_n \hat{\sigma}^{(n)} \vert \vert {}^8\text{Be}(18.15) \rangle \rvert^2.
\end{align}
Because there are no bounds on the proton coupling or neutron coupling individually, the axial vector coupling is only loosely constrained. 
Even taking into account the additional KTeV anomaly~\eqref{eq:KTeV} (the green bands in Fig.~\ref{fig:X17}b) still leaves a large part of the parameter space accessible. 

Unlike the beryllium decay, the required axial carbon matrix element, $\langle ^{12}\text{C}(\text{g.s.}) || \hat{D}_3^\sigma ||  ^{12}\text{C}(17.23) \rangle$, with $\hat{D}_3^\sigma$ the corresponding spin dipole operator, is not known \cite{Barducci:2022lqd}.
This relates back to the aforementioned tension.
We view the computation of $\langle \hat{D}^\sigma_3 \rangle$ as a pressing open problem.
Its resolution would immediately tell us whether one has a true discrepancy between the carbon and beryllium results---there would be no overlap between the carbon and beryllium measurements in the parameter space of the axial-vector couplings---or whether one has merely misinterpreted the data. Order-of-magnitude estimates suggest that the $^8$Be and $^{12}$C data could be harmonized using an axial vector X17 \cite{Barducci:2022lqd}.

Following the analyses in Refs.~\cite{Barducci:2022lqd, Hostert:2023tkg} we summarize the aforementioned constraints in Fig.~\ref{fig:X17} for the two most promising X17 scenarios, vector and axial-vector, with projected limits from Fig.~\ref{fig:reach} overlaid on top. Note that, unlike in Fig.~\ref{fig:reach}, here the shaded regions indicate allowed regions of the parameter space. Our ${}^8$Be limits differ slightly from Refs.~\cite{Barducci:2022lqd, Hostert:2023tkg} as we use a different mixing angle $\theta_{1^+}$. We have checked that when using their mixing angle we recover the their results. The actual value of the mixing angle has not been pinned down exactly (for example Refs.~\cite{Alves:2017avw, Feng:2016ysn, Kozaczuk:2016nma} use different values, see also the discussion in Ref.~\cite{Denton:2023gat}), meaning the available parameter space may be slightly larger than shown in Fig.~\ref{fig:X17}. For both the vector and axial-vector scenarios, changing the mixing angle effectively rotates the ${}^8$Be bands around the origin. Especially for the axial-vector scenario limits from $\gamma d \to e^+ e^- pn$ would constrain to what extent this rotation is permissible.  

In summary, if a calculation of the axial transition matrix element of $^{12}$C(17.23) to the ground state remains outstanding, constraints from $\gamma d \to e^+ e^- pn$ would be subdominant for a vector X17 scenario, but for what is at present the most likely X17 scenario, an axial vector, helpful to limit the available coupling range.

\section{Summary and outlook}\label{sec:outlook}
Low-mass dark sector searches are a very active area of research, fueled further by possible sightings of a new 17 MeV$/c^2$ boson, X17.
Compared to the electron and proton couplings, the coupling of a new boson in the 10--100 MeV$/c^2$ mass range to the neutron is poorly constrained.
To that end, we studied the efficacy of the deuteron photodisintegration process $\gamma d \to e^+ e^- pn$ at MAGIX@MESA in limiting the effective coupling of a new scalar, pseudoscalar, vector or axial-vector boson in the 10--100 MeV$/c^2$ mass range, using kinematics corresponding to quasi-free production on a neutron.
It turns out that for a scalar or vector scenario other methods yield better constraints on the neutron coupling.
However, for a pseudoscalar and axial vector scenario current bounds only limit the difference between the proton and neutron coupling. 
In this case, deuteron photodisintegration has the potential to yield limits directly on the neutron coupling.
Specifically, for the axial vector scenario our projections from $\gamma d \to e^+ e^-pn$ yield competitive constraints for the neutron coupling in light of X17.

Furthermore, besides providing exclusion limits on the neutron, via the reaction $\gamma d \to e^+ e^- pn$ one can also study the neutron polarizabilities, as has been previously done using $\gamma d \to \gamma pn$ \cite{Levchuk:1994ij, A2CollaborationatMAMI:2021vfy}. Such a study would be a worthwhile extension of the present work.

\section*{Acknowledgments}
The authors thank S. Schlimme, D. Barducci and C. Toni for helpful communications.
This work was supported by the Deutsche Forschungsgemeinschaft (DFG, German Research Foundation), in part through the Research Unit [Photon-photon interactions in the Standard Model and beyond, Projektnummer 458854507 - FOR 5327], and in part through the Cluster of Excellence [Precision Physics, Fundamental Interactions, and Structure of Matter] (PRISMA$^+$ EXC 2118/1) within the German Excellence Strategy (Project ID 39083149).

\appendix

\section{Calculation of the QED background}
\label{app:QED_BG}

\subsection{Conventions}
Our computation of the QED background starts with the QED vertex following from the QED interaction Lagrangian.
In our convention,
\begin{equation}
    \mathcal{L}_\text{QED} = - J_\mu A^\mu = e \bar{\psi} \gamma^\mu \psi A_\mu,
\end{equation}
The photon nucleon vertex $\Gamma^\mu_q$, is defined as
\begin{align}
    &\langle N(p') \, \vert \, J^\mu(0) \, \vert \, N(p)  \rangle \nonumber \\
    &= -e \bar{u}(p') \Gamma^\mu_q u(p) \nonumber \\
    &= -e \bar{u}(p') \left[ F_1(q^2) \gamma^\mu + \frac{i}{2 m_N} F_2(q^2) \sigma^{\mu\nu} q_\nu  \right] u(p), \label{eq:photonvertex}
\end{align}
where $N = p, n$ and $J^\mu(x)$ is the electromagnetic current operator, with $q = p' - p$.
For the Dirac and Pauli form factors, $F_1$ and $F_2$ respectively, we use the parametrization given in Ref.~\cite{Bradford:2006yz}. 

\subsection{Bethe-Heitler amplitude}
The tree-level diagram for the Bethe-Heitler process is shown in Fig.~\ref{fig:QED_diag}a.
The amplitude is given by
\begin{align}
    \mathcal{M}_\text{BH} &= \frac{ie^3}{(q - q')^2 } \varepsilon_\mu(\vb{q}, \lambda) L^{\mu\nu}_\text{BH} H^\text{BH}_{\nu},
\end{align}
where $\varepsilon^\mu$ is the photon polarization vector, with
\begin{align}
    L^{\mu\nu}_\text{BH} &= \bar{u}(p_-,s_-) 
    \bigg\{ \gamma^\mu  \left( \frac{\slashed{p}_- - \slashed{q} + m_e }{-2 p_- \cdot q} \right) \gamma^\nu \nonumber \\
    &\qquad + \gamma^\nu \left( \frac{ \slashed{q} - \slashed{p}_+ + m_e }{-2 p_+ \cdot q} \right)  \gamma^\mu  \bigg\} v(p_+, s_+),\\
    H^\nu_\text{BH} &=  \bar{u}(p_n,s_n) \Gamma_{q - q'}^\nu u(p_{\tilde{n}},s_{\tilde{n}}) .
\end{align}

\subsection{Born amplitude}
The tree-level Born amplitude is shown in Fig.~\ref{fig:QED_diag}b, and its amplitude is given by
\begin{equation}
    \mathcal{M}_\text{Born} = - \frac{ie^3}{q'^2} \varepsilon_\mu(\vb{q}, \lambda) L_\nu^\text{Born} H^{\mu\nu}_\text{Born},
\end{equation}
where
\begin{align}
    L^\nu_\text{Born} &= \bar{u}(p_-,s_-) \gamma^\nu v(p_+, s_+),\\
    H^{\mu\nu}_\text{Born} &= \bar{u}(p_n,s_n) \bigg\{ \Gamma^\nu_{-q'} \frac{( \slashed{p}_{\tilde{n}} + \slashed{q} + m_N )}{( p_{\tilde{n}} + q )^2 - m_N^2 } \Gamma^\mu_q \nonumber \\
    &\qquad + \Gamma^\mu_q \frac{( \slashed{p}_{\tilde{n}} - \slashed{q}' + m_N )}{( p_{\tilde{n}} - q' )^2 - m_N^2} \Gamma^\nu_{-q'} \bigg\} u(p_{\tilde{n}},s_{\tilde{n}}),
\end{align}
with $\Gamma^\mu_q$ given in Eq.~\eqref{eq:photonvertex}.

\subsection{Pion-pole amplitude}
For the pion-pole contribution we need the pion-photon and pion-nucleon vertices. 
The pion-photon vertex follows from
\begin{align}
    \mathcal{L}_{\pi^0 \gamma \gamma} &= \frac{e^2}{4} F_{\pi^0 \gamma \gamma} F^{\mu\nu} \tilde{F}_{\mu\nu} \pi^0,
\end{align}
where $\tilde{F}_{\mu\nu} = \tfrac{1}{2} \varepsilon_{\mu\nu\alpha\beta} F^{\alpha\beta}$ and $F_{\pi^0 \gamma \gamma}(0) = 1/(4 \pi^2 f_\pi)$, with $F^{\mu\nu}$ the photon field strength and $f_\pi \approx 92.4$ MeV the pion decay constant. 
Our convention is $\varepsilon_{0123} = +1$.
The photon-nucleon vertex follows from
\begin{equation} \label{eq:NNg}
    \mathcal{L}_{\pi N N} = g_{\pi N N} \bar{N} \pi^a \tau_a i \gamma_5 N,
\end{equation}
where $\tau^a$ are the Pauli matrices, $N = (p, n)$ and $g_{\pi N N} = g_A m_N / f_\pi$, with the axial-vector coupling from nuclear $\beta$-decay $g_A = 1.27$.
The pion-pole amplitude is shown in Fig.~\ref{fig:QED_diag}c, and is given by
\begin{equation}
    \mathcal{M}_{\pi^0} = - F_{\pi^0 \gamma \gamma} g_{\pi N N}\frac{i e^3}{q'^2} \varepsilon^\mu(\vb{q},\lambda) L^\nu_{\pi^0} H^{\pi^0}_{\mu\nu} ,
\end{equation}
where
\begin{align}
    L^\nu_{\pi^0} &= \bar{u}(p_-, s_-) \gamma^\nu v(p_+, s_+), \\
    H_{\mu\nu}^{\pi^0} &=  \frac{i}{(q-q')^2 - m_\pi^2} \varepsilon_{\mu\nu\alpha\beta} q^\alpha q'^\beta \bar{u}(p_n, s_n) \gamma_5 u(p_{\tilde{n}}, s_{\tilde{n}}).
\end{align}

\subsection{Non-Born neutron polarizability amplitude}
Lastly, we account for nucleon polarizability contributions following the formalism in Ref.~\cite{Lensky:2017bwi}.
In short, one may parameterize the unpolarized Compton process, $N(k) \gamma(q) \to N(k')\gamma(q')$, often subtracting the Born and pion exchange contribution, with 18 independent tensors $T^{\mu\nu}_i$ and their corresponding coefficients $B_i(q^2, q'^2, q \cdot q', q \cdot P)$, where $P = \tfrac{1}{2}(k + k')$. 
For unpolarized real Compton scattering only two tensors contribute (see Appendix A in Ref.~\cite{Lensky:2017bwi}),
\begin{align}
    M^{\mu\nu}_\text{RCS, spin avg.} &= B_1(0, 0, q \cdot q', q \cdot P) T_1^{\mu\nu} \nonumber \\
    &\qquad + B_2(0, 0, q \cdot q', q \cdot P) T_2^{\mu\nu},
\end{align}
where
\begin{align}
    T_1^{\mu\nu} &= - (q \cdot q') g^{\mu\nu} + q'^\mu q^\nu,\\
    T_2^{\mu\nu} &= (2 m_N \nu)^2 \left( -g^{\mu\nu} + \frac{q'^\mu q^\nu}{k \cdot q'} \right) \\
    &\qquad - 4 (q \cdot q') \left(P^\mu - \frac{q \cdot P}{q \cdot q'} q'^\mu \right) \left(P^\nu - \frac{q \cdot P}{q \cdot q'} q^\nu \right),
\end{align}
with $\nu = (q \cdot P) / m_N$. 
The low-energy expansions for $B_1$ and $B_2$ are given by
\begin{align}
    B_1(0, 0, q \cdot q', q \cdot P) &= \frac{1}{\alpha} \beta_{M1},\\
     B_2(0, 0, q \cdot q', q \cdot P) &= - \frac{1}{\alpha} \frac{1}{(2 m_N)^2} (\alpha_{E1} + \beta_{M1}),
\end{align}
where the values for the neutron polarizabilities are \cite{ParticleDataGroup:2022pth}
\begin{align*}
    \beta^n_{M1} &= (3.7 \pm 1.2) \times 10^{-4} \text{ fm}^3, \\
    \alpha_{E1}^n + \beta^n_{M1} &= (15.2 \pm 0.5) \times 10^{-4} \text{ fm}^3.
\end{align*}

The polarizability amplitude is illustrated in Fig.~\ref{fig:QED_diag}d and given by
\begin{equation}
    \mathcal{M}_\text{Pol.} = -\frac{ie^3}{q'^2} \varepsilon^\mu(\vb{q},\lambda) L^\nu_{\text{Pol.}} H^{\text{Pol.}}_{\mu\nu} ,
\end{equation}
with
\begin{align}
    L^\nu_{\text{Pol.}} &= \bar{u}(p_-, s_-) \gamma^\nu v(p_+, s_+) \\
    H_{\mu\nu}^{\text{Pol.}} &= \bar{u}(p_n, s_n) M^\text{RCS, spin avg.}_{\mu\nu} u(p_{\tilde{n}}, s_{\tilde{n}}).
\end{align}
where the blob in Fig.~\ref{fig:PWIA} indicates the polarizability contribution.
Note that in our case $\sqrt{q'^2} \ll 1$ GeV, so using the real Compton amplitude in place of the single-virtual Compton amplitude is a very good approximation.

\section{Verification of the QED background}
\label{app:QED_compare}

\begin{figure}[tb]
    \centering
    \includegraphics[width=8.6cm]{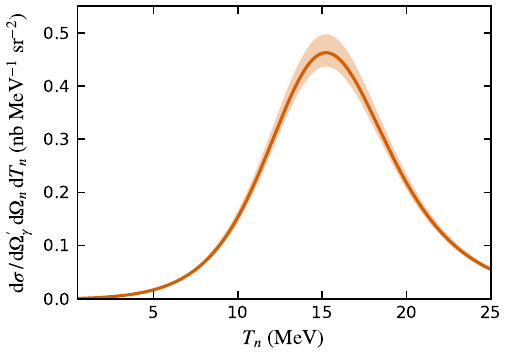}
    \caption{The in-plane differential cross section of $\gamma d \to \gamma pn$ as a function of the neutron kinetic energy, $T_n$. We fix $E_\gamma = 100$ MeV, $\theta'_\gamma = 135\text{\textdegree}$ and $\theta_n = -20\text{\textdegree}$.}
    \label{fig:Levchuk}
\end{figure}

In general, there is no calculation of the QED background of $\gamma d \to e^+ e^- pn$ that we can directly compare to. However, there has been a calculation of the QED background in $\gamma d \to \gamma pn$ \cite{Levchuk:1994ij}. Up to phase-space factors this is, in principle, identical to our calculation with the Bethe-Heitler contribution removed. Therefore, it can serve as a useful intermediate cross check. Of course, the methods used in Ref.~\cite{Levchuk:1994ij} are not identical to our methods, so we cannot expect a completely identical result. Nevertheless, our results should agree to at least the same order of magnitude.

To that end, we have repeated the PWIA calculation of Ref.~\cite{Levchuk:1994ij}. Using their notation, we have computed the in-plane differential cross section $\mathrm{d} \sigma / \mathrm{d}\Omega_{\gamma'} \mathrm{d}\Omega_{n} T_n$ as a function of $T_n$, where $T_n$ is the kinetic energy of the outgoing neutron and $\Omega_{\gamma'}$ and $\Omega_{n}$ are the solid angles of the outgoing photon and neutron, respectively (see Ref.~\cite{Levchuk:1994ij} for details of the calculation). This is shown in Fig.~\ref{fig:Levchuk} and should be compared to the dotted line in Fig.~5a of Ref.~\cite{Levchuk:1994ij}. As can be seen, both calculations agree, giving a useful cross check of our QED background.

\section{Dependence reach of neutron coupling on integration range}\label{app:reach_int_range}

\begin{figure}[bt]
    \centering 
    \includegraphics[width=8.6cm]{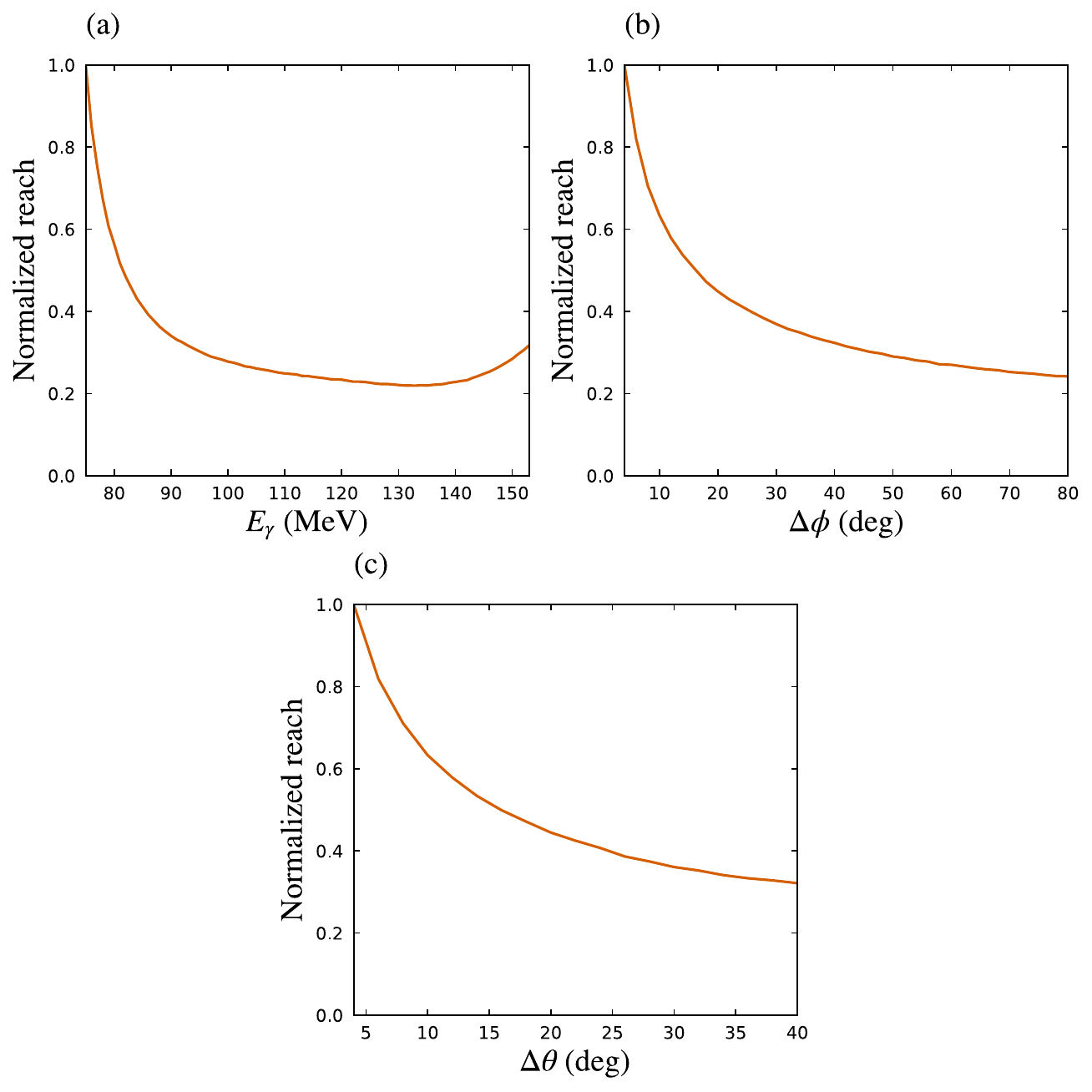}
    \caption{The dependence of the reach on the size of the integrated phase space for a vector-like $X$ normalized to its value at the start of the $x$-axis. Here we vary (a) $E_\gamma$, (b) $\phi_\pm$ or (c) $\theta_\pm$ while keeping the rest of the kinematic variables fixed to detector setting 2 (see text). We set $m_X = 65$ MeV$/c^2$ and $\delta m_X = 0.1$ MeV$/c^2$.}
    \label{fig:phase_reach}
\end{figure}

From Eq.~\ref{eq:reach_explicit} it is clear that the reach of the coupling of $X$ to the neutron depends on the size of the phase space we integrate over.
Given that the detector settings in Table~\ref{tab:detector} are estimates, it is important to see to what degree changing the accessible phase space influences the reach. 
The ranges of $|\mathbf{p}_\pm|$ are unlikely to change in the final MAGIX@MESA setup, and cutting the range of $\theta_n$ has little influence on the reach (see Sec.~\ref{sec:reach}).
Therefore, we focus on the influence of $\theta_\pm$, $\phi_\pm$ and $E_\gamma$.

The general trend is that increasing the available phase space improves the reach, albeit with diminishing returns.
This is illustrated in Fig.~\ref{fig:phase_reach}. 
Here, we take detector setting 2 (with $m_X = 65$ MeV$/c^2$ and $\delta m_X = 0.1$ MeV$/c^2$) and either increase the integration range of $\phi_\pm$ or $\theta_\pm$ symmetrically, 
\begin{align*}
    \phi_+ &\in [180 \text{\textdegree} - \Delta \phi_\pm, \ 180 \text{\textdegree} + \Delta \phi_\pm] ,\\
    \phi_- &\in [-\Delta \phi_\pm, \ \Delta \phi_\pm ],   
\end{align*}
or
\begin{equation*}
    \theta_\pm \in [120\text{\textdegree} - \Delta \theta_\pm, \ 120\text{\textdegree} + \Delta \theta_\pm],
\end{equation*}
respectively, keeping the other parameters fixed. We have assumed a vector-like scenario, however we have checked that the same conclusions hold for the other parity assignments.

In Fig.~\ref{fig:phase_reach}a we see that, as a function of $E_\gamma$, the reach has an optimum. The reason here is that shifting the energy roughly corresponds to translating the exclusion limits in Fig.~\ref{fig:reach}---at a certain point one hits the kinematic boundaries.

In Fig.~\ref{fig:phase_reach}b we see that increasing the integration range to be more out of plane quickly increases the reach, before flattening out.
A similar trend is visible in Fig.~\ref{fig:phase_reach}c, when one increases the range of the polar angles.
So, if one integrates over a very small part of the phase space, it is worthwhile to try to access more.
If one already can integrate over a large area of phase space, increasing the phase space further is not as useful.

\bibliography{bib.bib}

\end{document}